# An Evolutionary Perspective on the Design of Neuromorphic Shape Filters

**Ernest Greene[1], Member, IEEE**
[1]Laboratory for Neurometric Research, Department of Psychology
University of Southern California, Los Angeles, CA 90089 USA

**ABSTRACT** A substantial amount of time and energy has been invested to develop machine vision using connectionist (neural network) principles. Most of that work has been inspired by theories advanced by neuroscientists and behaviorists for how cortical systems store stimulus information. Those theories call for information flow through connections among several neuron populations, with the initial connections being random (or at least non-functional). Then the strength or location of connections are modified through training trials to achieve an effective output, such as the ability to identify an object. Those theories ignored the fact that animals that have no cortex, e.g., fish, can demonstrate visual skills that outpace the best neural network models. Neural circuits that allow for immediate effective vision and quick learning have been preprogrammed by hundreds of millions of years of evolution and the visual skills are available shortly after hatching. Cortical systems may be providing advanced image processing, but most likely are using design principles that had been proven effective in simpler systems. The present article provides a brief overview of retinal and cortical mechanisms for registering shape information, with the hope that it might contribute to the design of shape-encoding circuits that more closely match the mechanisms of biological vision.

## I. INTRODUCTION

The computational skills of the human brain are a wonder, so it is easy to understand why many research engineers are interested in developing neuromorphic circuits, i.e., electronic implementation of neuron mechanisms. We are all impressed by the ability of the human brain to register and store vast quantities of visual information. But what is sometimes missed is an awareness of the degree to which each mechanism has been tailored over many millions of years – or hundreds of millions -- to be near optimal for achieving survival of intervening species. It is understood that the retina has anatomical and physiological filters that can effectively encode image information, but often the functioning of visual cortex is seen as a tabula rasa.

Ethologists can readily affirm that the inborn visual skills of many species are exceptional from the start. A newborn gazelle can be up and running with its mother within an hour. The anatomy and physiology of its visual cortex are already sufficient to mediate perception of objects, depth, and motion, as evidenced by the effectiveness of its behavior. Experimental study of the anatomy and physiology of cortical systems affirms the pre-programmed complexity, some of which will be discussed subsequently. The relevance, at this point, is to convey my belief that the most common approach to neural network design has been a mistake.

I will not be discussing how motion, color, texture, and brightness gradients contribute to the analysis of image content. The immediate focus will be on how contours, the lines and edges of a given object, make it possible to identify an object from a line drawing as well as from a photograph, as illustrated in Figure 1. Further, can the visual system accomplish this recognition if the object has a novel shape and has been seen only once, so the identification is not based on long-term memory? What system design provides for recognition of the novel shape if it is subsequently displayed at a different location within one's visual field, or at a different orientation, or a different size? Can it be identified if the boundary has been fragmented, as might occur when an object is seen behind branches and leaves? The human visual system is fully capable of successful identification of objects under all of these conditions. I will make the case that none of these visual skills are unique to the human brain, or even the brains of mammals. Rather, these mechanisms are available to most vertebrate species, perhaps to all, having been tailored by evolution over hundreds of millions of years.

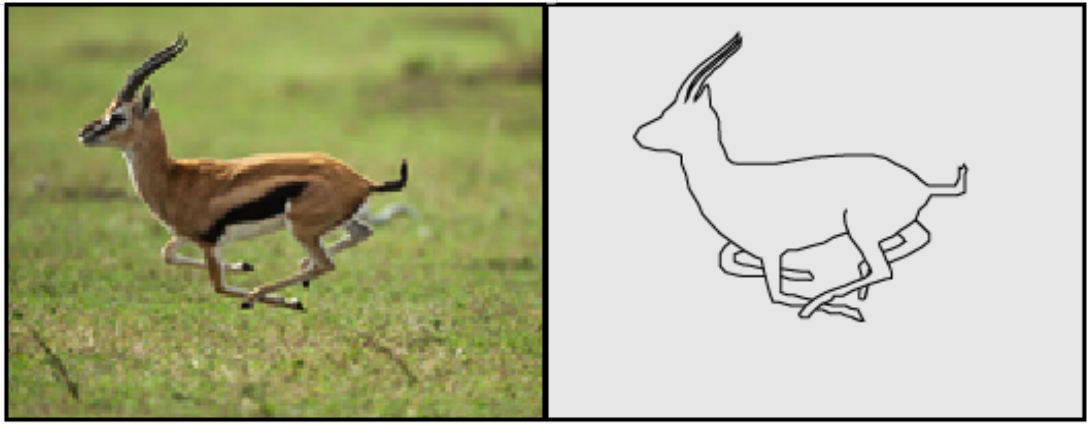

**FIGURE 1.** Effective neuromorphic image analysis must at least be able to identify objects using only the outline boundary. A number of boundary descriptors have been developed on the assumption that shape recognition is based on contour attributes. Biological vision may have evolved using different principles.

In this article I will describe various findings that argue for structured filter operations in the retina and with the cortical filters being equally well structured. Where the design principles for an advanced visual skill are still unknown, such as recognition of objects, I will assert that the mechanism is not based on modifying synaptic connectivity in successive neuron populations through countless training trials.

## II. Retinal Filters for Marking Contrast

It is well known that image information in vertebrates is transferred from photoreceptors, through bipolar cells, to ganglion cells, which are the source of optic nerve fibers.[1],[2] For the ganglion cells that are pertinent here, each has a localized "receptive field" -- a term that specifies the area that will respond to a light stimulus. The receptive field will usually consist of two zones, a central region and a surrounding region (an annulus), each responding to light in opposite ways. This is commonly described as a "center/surround" design.[3],[4]

The center corresponds to the area covered by the branches of ganglion cell dendrites. The location of a given ganglion cell determines the size of the central area, i.e., the number of photoreceptors that provide it with stimulation. In the fovea of primates and many other species there is a ganglion cell for each receptor but in peripheral retina a ganglion cell may receive stimulation from a pool comprised of hundreds of photoreceptors. The opposite-acting surround influence is delivered through horizontal or amacrine cells that have received their activation from bipolar cells.[5],[6] Whether the surround influence is delivered by horizontal cells, amacrine cells, or both, is still somewhat unsettled. This may not be critical with respect to functionality.

Perhaps it is a bit less well known that some ganglion cells signal an increase in light that falls at the center of their receptive fields, whereas others signal a decrease in light.[7] This dual code is of sufficient importance to warrant additional discussion of how it is generated. To do so, we need to move back up the chain and describe how light is registered by the photoreceptors, conveyed to bipolar cells, and from there to the ganglion cells described above.

The photoreceptors of all vertebrates, including fish, amphibia, reptiles, birds, and mammals, respond to a decrease of light by an increase in membrane potential (depolarization), and they hyperpolarize if the light level is increased.[8],[9] Figure 2 shows that each photoreceptor makes synaptic contact with two bipolar cells, one of the bipolar cells receiving a signal that matches the polarity change of the photoreceptor and the other that reverses the polarity.[10]-[12] These are called "ionotropic" and "metabotropic" synapses, respectively. Details for how the signal is reversed are not needed, our focus is on the functional outcome.

We now have the basis for naming the responses produced by transitions of light. If the light is decreased, the photoreceptor depolarizes, which is passed through the ionotropic synapse as a depolarization, so we can describe that bipolar neuron as carrying an "OFF" signal. Conversely, if the light is increased, the photoreceptor hyperpolarizes, which depolarizes the other bipolar neuron through the metabotropic synapse. We can describe this neuron as providing an "ON" signal. Figure 2 completes the basic description of ON and OFF information channels by showing that each type (class) of bipolar cell will selectively connect to ganglion cells, or more precisely, to the center of the receptive field of a corresponding ganglion cell. ON ganglion cells receive input from ON bipolar cells, and OFF ganglion cells receive input from OFF bipolar cells.[14]-[16]

Note that interactions among most of the retinal neurons, up to the point of ganglion-cell firing, are accomplished by "graded" (analog) changes in membrane potential, transferred from one cell to another through chemical and electrical synapses. There are some examples of miniature action potentials and one class of amacrine cell that radiates spikes through an arbor of axons. But most of the signal transmission is analog, an approach favored by a number of modelers and research engineers doing neuromorphic retinal image encoding.

The centers of ON ganglion cells "tile" the retina, with the centers being positioned edge-to-edge, similar to a tight array of coins on a table-top.[17],[18] The OFF ganglion cells also tile the retina, with overlap of ON and OFF fields being accomplished while maintaining functional separation of response through selective micro-connectivity of synapses.[13] Thus for a given location on the retina, an increase in light level with be signaled through ON ganglion cells and a decrease at that same location will be signaled through OFF ganglion cells.

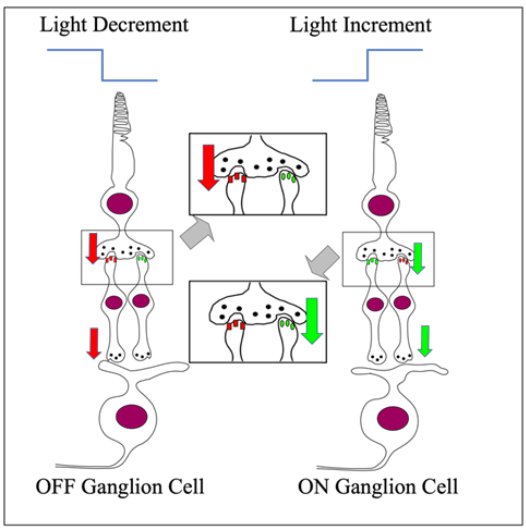

**FIGURE 2.** Retinal filters register local brightness differentials (contrast) using center/surround receptive fields. The sources of input to the centers is illustrated here. A field that registers an increment of light to the center but not the surround will activate an ON-ganglion cell and a field that registers a decrement will activate an OFF-ganglion cell. This two-channel design is made possible by ionotropic synapses that directly transfer the response of photoreceptors to a decrease in light, and metabotropic synapses that invert the photoreceptor response to an increase in light.

What function would be provided by the oppositional center/surround design? Figure 3 shows stimuli that are localized on receptive field centers, one that is brighter than background on the left and one that is darker than background on the right. The differential in brightness between the central area and the background, i.e., the contrast, will activate the ON and OFF ganglion cells, respectively. However, the same ganglion cells will not register a uniform stimulus because the stimulation of the center is counterbalanced by stimulation of the surround. This will be the case whether the uniform stimulation is bright or dark. The ganglion cell is designed to register when there is a localized departure from a uniform background.

It is likely that a relatively primitive vertebrate visual system found it beneficial to register the presence of objects against a uniform background. For an ancestral fish looking up toward the surface of the water, an OFF filter could register a small dark object against the bright background. Any filter stimulated only by light from a zone adjacent to the object would not be generating any signal, as activation of the filter's center would be cancelled by activation of its surround. This would be the case for ON as well as OFF ganglion cells, and also, irrespective of the overall brightness of the background. The center/surround design provides a filter mechanism for registering the contrast of a localized region of the visual field, this serving to detect objects while ignoring background.

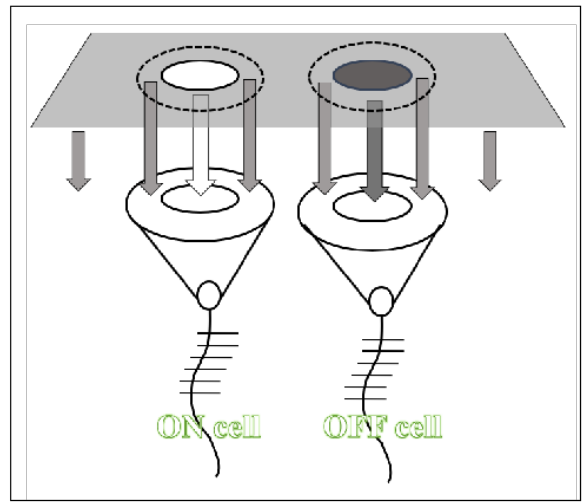

**FIGURE 3.** The figure shows how ON and OFF ganglion cells register local contrast differentials in their receptive fields. The ON cell will fire if its center is receiving more light than the surrounding (background) region, and the OFF cell will fire if its center is receiving less light than the surrounding (background) region.

Figure 4 illustrates how an array of center/surround receptive fields would respond for a silhouette that was larger than the size of each receptive field. One can see that both the interior and the exterior of the object would provide relatively uniform stimulation of a given receptive field. Activation of the center would be counterbalanced by an equal activation of the surround, so neither would be generating any signal. But receptive fields that happened to lie at the edge of the object would detect a contrast differential, given that the center could be fully activated whereas the surround of the cell would be only partially activated. Simply put, the collective response of an array of center/surround ganglion cells could "mark" locations around the boundary of the object, thus providing a potential source of information about the object's shape.

The example given above can be reversed, with the primitive fish looking down to see an object that is lighted from above. The object might be bright against a dark background and interior contours might be visible. We can still invoke the basic concepts, reversed with respect to the signaling of ON and OFF ganglion cells, and concede that the discussion of shape recognition becomes more complicated when interior contours are considered.

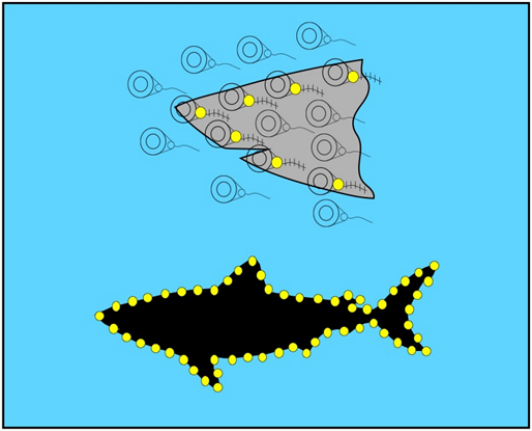

**FIGURE 4.** Differences in the amount of activation of center versus surround can mark the edges of a shape. A center that lies just inside the edge of this dark figure receives less light than the surround, which gets some light from the background water. The array of these edge markers provides elementary shape cues. Fields that are stimulated only by the background water register no net differential in brightness, and thus are silent. The same is true for receptive fields that receive stimulation from the interior of the object.

However, for present purposes it is sufficient to limit discussion to contour markers that fall on the boundary of the object. Therefore, we move forward on that basis.

The concept outlined in Figure 4 is not novel; the use of center/surround opposition for edge detection and edge enhancement has been offered countless times since the initial discovery of this design.[3] Further, as we will see shortly, means to register lines and edges (contours) has been carried forth into discussion of cortical design. However, providing this evolutionary framework for how the mechanism contributes to survival may prove useful for subsequent discussion of functional goals. Please keep the imagery in mind as we will be returning to this issue.

## III. Cortical Filters for Registering Lines and Edges

The filter functions of neurons in primary visual cortex of mammals (V1), and especially those that have been demonstrated for primates, provide the gold standard for discussing models of human visual function. Since their initial discovery by Hubel & Wiesel more than half a century ago,[19] the image processing done by orientation selective neurons has been viewed as an indispensable component of shape encoding and an essential first step toward shape identification.

Anatomical and physiological evidence supports the proposition that the responses of orientation-selective neurons are driven by the output from short arrays of retinal ganglion cells.[20] We need not be concerned about details for how the signals are relayed through thalamus and through layer 4 of V1 before converging onto orientation-selective neurons. However, it is pertinent that, whatever the species, rat, cat, or human, the means for connecting the ganglion cell array to a given orientation-selective neuron has been preprogrammed by the animal's genes.[21] The filter properties of the receptive field of a given orientation-selective neuron have been put there by anatomical convergence of axons arising from the ganglion cells, passing through several relay stages, and arriving to make synaptic contact with the cell. It is not programmed by countless training trials that begin with random connectivity that must alter linkage or synaptic strength to make it responsive to only one short set of aligned retinal filters. It has been programmed by hundreds of millions of years of evolutionary pressure, providing an overall system-design that has proven to be extremely successful in securing survival of ancestral mammals.

Figure 5 illustrates the structural precision that evolution has achieved. The left panel shows a composite image that was derived from optically monitoring activation of V1 in monkey in response to moving bars.[22] The right panel shows an idealized version of the tiling. Activation of cortical neurons produces small changes in opacity of the tissue that can be registered by a camera, and the changes can be seen across the many locations that have been simultaneously stimulated by a moving bar. If vertical bars are being passed across the display, one will see a patchwork of activation being registered by the camera as the reflectance of each zone is briefly altered. One can plot those locations into a recorded image as a specific color, e.g., red, and use other colors to designate the locations that will be activated when the bars pass across the screen at other orientations. The resulting composite shown in Figure 5 shows a radial structure for small groups of orientation-selective neurons that map the input from small patches in the retina. Each rainbow swirl of color is designated as a "pinwheel," likely because it is suggestive of colors on a child's top. (There is a long history of earlier work done with extracellular recording that uses the term "hypercolumn" to describe how these neurons are organized. That record is not especially relevant to the current discourse.) One can see that V1 is tiled with pinwheels that register contrast differentials, responding in particular to the lines and edges that are present in a given image.

## IV. Elementary Shape Filters

One might think the complex cortical filters would be an essential starting point for registering the lines and edges that define a given shape But the ancestral fish, mentioned above, would have needed elementary shape filters to survive and pass on its visual skills to the many billions of progeny that followed. Certainly modern fish provide examples of how elementary shape filters contribute to survival, e.g., deciding whether an object is a predator.[23],[26] Activation of center/surround filters is not sufficient to determine whether the observing fish should turn and flee, or swim toward the object on the chance that it is a species that could be dinner. The fish needs elementary shape filters that can register (summarize) the pattern of marked locations and provided a basis for choosing a beneficial action.

The shape filter that allows for the animal's survival cannot be specific for a given size, for a predator must be

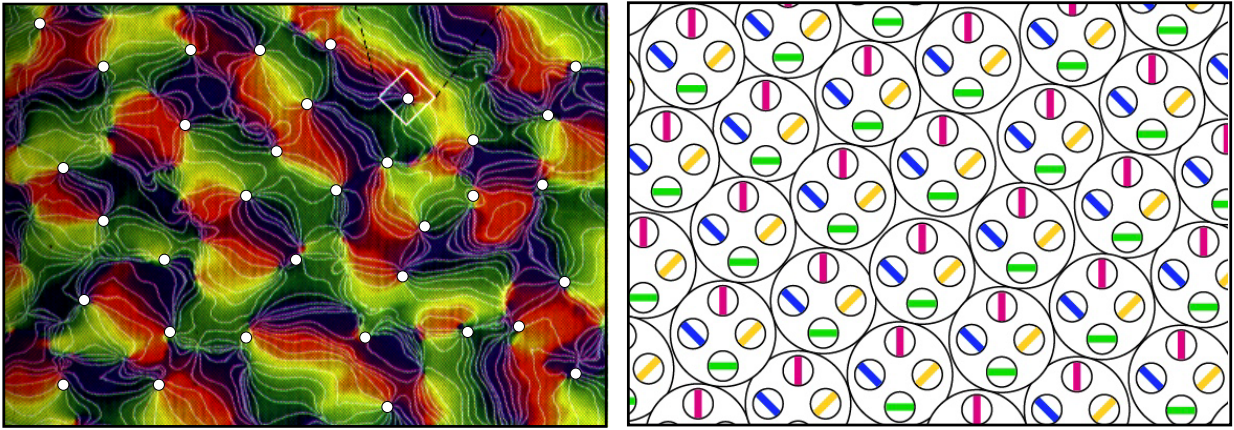

**FIGURE 5.** The left panel provides a pseudocolor image of the surface of V1 in Macaque, where each color represents activation of orientation-selective neurons that were stimulated by moving bars. White dots have been added to the image, showing the center of each "pinwheel," this being a radial configuration of the neurons responding to the various orientations. The right panel shows an idealized diagram of pinwheel tiling of the cortex.

spotted at various distances.[27]-[29] If the object has come closer, size invariance is needed to assure recognition after marker locations have changed. One glimpse may not provide sufficient information to make a decision about what was seen. At the next moment the object or the observer may have moved, so the summary that was derived from one set of boundary-marking filters needs to be matched with what another set provides. In other words, the mechanism needs to be translation (position) invariant.[23],[30] Movement may have altered the orientation of the initial record, so effective identification of the shape requires rotation invariance.[31],[32]

Further, the entire complement of boundary markers for a given shape may not all be present due to occlusion. If the object is hidden behind a dense thicket of sea-plants or coral, only fragmented portions of the boundary may be visible at a given moment (see Figure 6). Therefore, the filter must provide a summary that is robust, allowing identification of the object using a minimal set of boundary markers.[33]

It is entirely within the capacity of genes to pre-wire inborn shape filters. If artificial neural networks can provide for invariant discrimination of shapes with less than 100,000 training trails, the requisite connectivity could be selected by evolution over hundreds of millions of years.

However, even for shapes that are learned, the core mechanism for learning a given marker pattern would likely provide for quick encoding and storage. Those who study fish behavior can affirm the ability to fish to spot a dangerous predator after only an initial brief encounter. A naive young fish might escape a first attack but would not likely have many opportunities to be that lucky. To not provide for this filter capacity is to assure a high probability of failure to escape on the second or third encounter. By the millionth generation, or the hundredth millionth, a way to quickly register, summarize, and store the shape of a predator would surely have evolved.

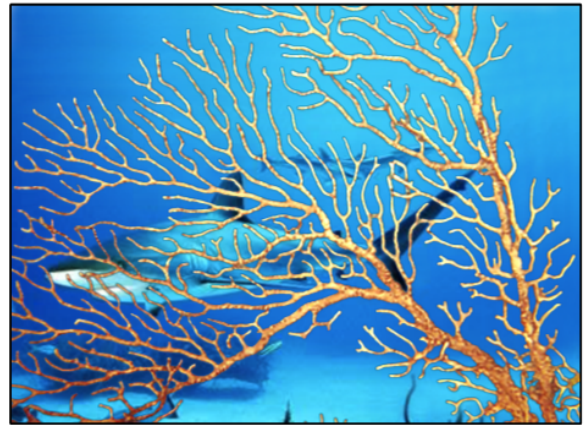

**FIGURE 6.** Biological vision can identify shapes even when occlusion provides only a partial view of the boundary. An effective neuromorphic shape filter should be able to identify shapes even when a reduced number of markers is provided.

The visual skills that evolved in fish allow for identification of appropriate prey, members of one's own species, effective navigation of underwater terrain, and such.[30] Any shape-filter operations that were not provided during incubation must be quickly manifested and fine-tuned shortly after hatching. The development of elementary shape filters cannot require numerous training trails, for a single bad choice can be lethal. There would be an evolutionary premium on developing a mechanism for one-trial learning.

It is relevant to note that the visual skills of modern fish are provided by two key structures – retina and optic tectum. The optic tectum is a homolog of the superior colliculus in mammals. In mammals, its major function is thought to be for control of eye movements – reflexive saccades and as a relay for voluntary saccades. There is minimal evidence, perhaps due to lack of investigational effort, of shape filtering by the superior colliculus of mammals. The common thought is that shape analysis is relatively rudimentary in non-mammalian vertebrates and the cortex of mammals has provided new and improved shape-recognition tools.

## V. Cortical Shape Filters

There is clinical evidence for rudimentary perception that might be based on residual function of the superior colliculus, often described as "blindsight." With extensive damage to primary visual cortex, a patient suffers dramatic loss of the ability to see in large portions of the visual field. He or she will report being totally blind if all of this cortical area is removed or degenerates from stroke, anoxia, or other sources of tissue injury. But over fifty years ago Lawrence Weiskrantz reported that these patients did retain the ability to register some kinds of visual stimulation. [34],[35]

Notwithstanding the blindsight findings, clearly the patients with damage to V1 are not able to identify specific shapes as might be needed for reliable navigation through an unfamiliar room, reaching to grasp a fork rather than a spoon, or reporting whether the silhouette of a bird or a goat was displayed. As described above, the visual system of a modern fish would likely be able to accomplish this level of shape discrimination. So, it is possible that the elementary shape filters that were available to non-mammalian computer scientists, and research engineers. It is a vapid concept, for it assumes that the shape filters must be developed anew for each newborn. The visual skills of pre-mammalian species across millions of years of evolution are viewed as being too primitive. Instead, one must tailor the new shape-recognition skills through trial-and-error encounter with the external world to achieve the remarkable levels that humans can manifest. I find it strange that so many have embraced this position.

So, let's provide an overview of some basic skills that cortical filters should provide if they are to match the elementary shape filters of non-mammalian vertebrates. The first panel in Figure 7 shows an object rendered with various brightness levels, colors, textures, and edges. In the second panel one can see that everything except the fine-line outline of the object's boundary can be eliminated and the object can still be recognized. The third panel shows that it can be identified even if one uses discrete dots as boundary markers.

Research has shown that a great many real-world objects can be named even if the complement of boundary markers is exceptionally sparse. Figure 8 provides a few examples from an experiment that asked subjects to name shapes based on sparse displays of boundary markers.[36] The figure illustrates the finding that very few boundary markers were vertebrates have been rendered mostly non-operational, perhaps even eliminated from the cellular machinery of the human brain. Alternatively, these shape filters may still be in use as a complement to cortical mechanisms but are lost if primary visual cortex is damaged.

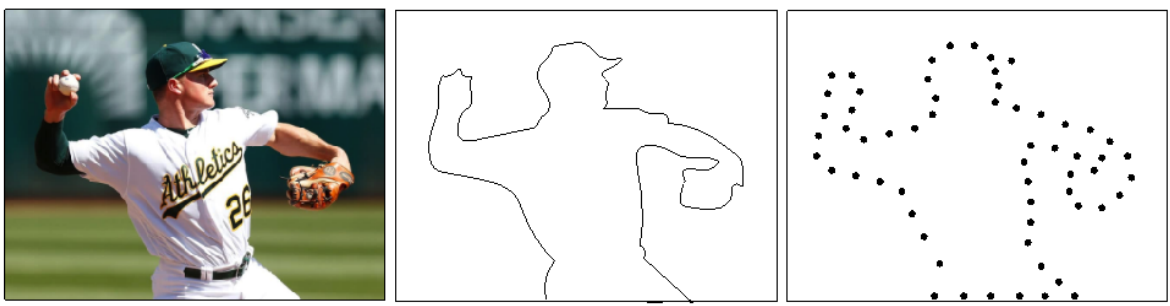

**FIGURE 7.** Not only can the visual system identify shapes that are represented using only the outline boundary of the shape, a sparse array of dots can provide for recognition. Research from my laboratory has demonstrated that a large range of objects, e.g., animals, plants, vehicles, tools, furniture, can be identified from spaced boundary markers. [36]

Whatever the case, it seems clear that primary visual cortex and its connections to other occipital, temporal, and parietal areas are providing most of the image encoding in the human brain, with the ability to register and summarize shapes being most relevant here. How shall we design these new and improved cortical shape filters? How about taking the output of orientation-selective neurons, i.e., those found in the pinwheels of V1 as described above, and randomly distributing the connections to occipital and inferotemporal neurons? Then we require many thousands of training trials to derive a shape-selective response, or many tens of thousands if one wants translation, rotation, and size invariance. This has been the approach for neural network modeling, as implemented by countless neuroscientists, required for retrieving the relevant memory, i.e., for object recognition. This affirms that the shape summary for a sparse pattern is consonant with the summary provided by the full boundary.

We also noted that the elementary shape filters need to generate a summary without requiring numerous displays of the shape, to better ensure survival and reproduction of the animal. Optimally, a summary should be generated with a single display, which can be described as one-trial learning.

Another set of experiments that called for "match recognition" of shapes are diagrammed in Figure 9.[37] Each trial of the task briefly displayed a target shape, followed quickly by a comparison shape that provided some of the target's boundary markers, or showed markers derived from a non-target shape. All target and non-target shapes were unknown, meaning that each was constructed as an arbitrary set of curves and straight segments, providing an outline boundary that did not resemble any known object. A given target shape was displayed only once to put the focus

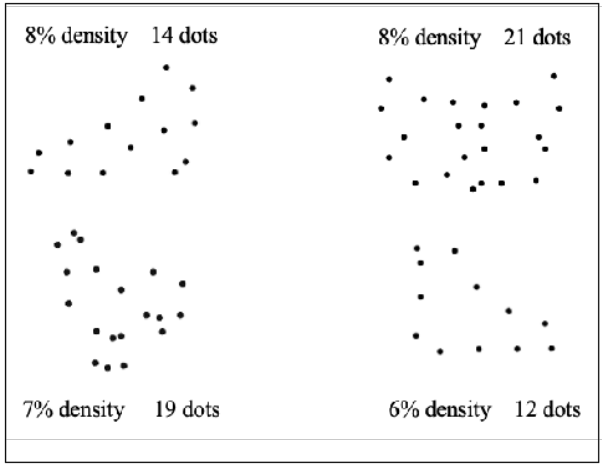

**FIGURE 8.** Human subjects were asked to name objects that were displayed as lighted dots in an LED display. At 100% density the dots would be adjacent, and lower densities provided a gap between the dots. The figure specifies the mean number of dots (and density) at which each shape was correctly identified.

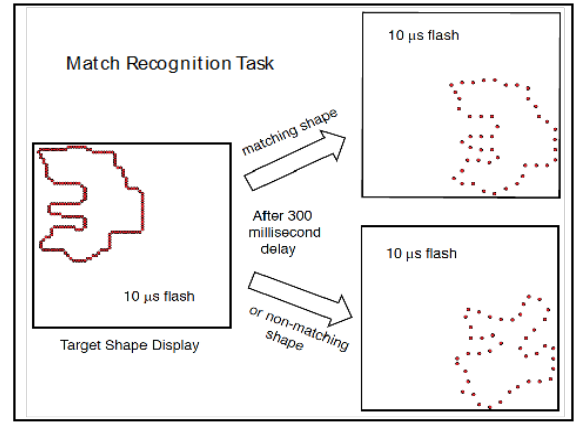

**FIGURE 9.** Amorphous shapes not resembling any known shape were briefly displayed as targets, each only once. A moment later a comparison shape was displayed, which was either a low-density version of the target or a low-density version of a different shape. Subjects were able to correctly choose whether the comparison shape was the "same" or "different" from the target. Decisions were well above chance even when the comparison shapes had very low density. Translation invariance is illustrated, but size and rotation invariance were also found.

on the shape-encoding process by precluding any learning.

The comparison shape was shown 300 milliseconds after display of the target shape and the subject would typically respond within 2-3 seconds. Subjects were able to say whether the comparison shape was the same or different from the target with a probability that was well above chance and did so even when the quantity of markers provided in the comparison shape was relatively small. Match-recognition was well above chance when the comparison shape was displayed at a different location than the target, or with a different size, or when it was rotated. In other words, the shape filter mechanism provided summaries that allowed comparisons that were translation, size, and rotation invariant.

Overall, it seems clear that cortical filters serve to encode shapes, i.e., provide shape summaries, with immediacy that has contributed to the survival of mammalian species. None of these operations seem far removed from what non-cortical mechanisms can do. Those visual skills appear to be available to fish, using the neural machinery of the retina and optic tectum.[23]-[33] These skills have substantial benefit, serving to increase the chances that a newborn (or newly hatched) animal will survive. It seems unlikely that they would be totally abandoned as new cortical tools were evolving.

## VI. Global Shape Filters

I submit that there has been far too much emphasis on the local contour attributes that are registered by orientation-selective neurons. The left panel of Figure 10 shows a perfect circle formed by a thin line. A complement of orientation-selective neurons would be activated, each registering a local portion of the line, which can be designated as a line segment. The location of the segment within the circle determines which cortical pinwheel is stimulated, and the orientation of the segment determines which neuron(s) within the pinwheel will fire. The focus of much theory is on the response properties of these cortical neurons. Orientation of the segment is considered to be a critical piece of shape information, as reflected in the use of that attribute in naming the neurons. Continuity of the contour is assumed to be very relevant, given the elongated excitatory region within the neuron's receptive field. There is some evidence for curvature being a factor in which neuron will be activated and some modeling of V1 neurons includes curvature as a shape attribute.[38]

However, though one might insist that only the first panel of Figure 10 displays a proper circle, the second and third panels demonstrate that orientation, curvature, and continuity of contours are not essential for perceiving circularity. The second panel uses line segments that lie at different orientations, lack curvature, and are disconnected. The third panel eliminates line segments altogether, providing only a pattern of disconnected dots. Yet each of these configurations can be characterized as being circular. The basis for this perception was described by the Gestalt School of Psychology more than a century ago.[39] It is the "global" attribute(s) of each configuration that makes it circular. Gestalt mechanisms are often cited with respect to shape perception, but almost always as a concession to the lack of specifics or insight about how the perception is being generated.

One might note that neurons in layer 4 of a V1 pinwheel register signals delivered from individual retinal ganglion cells and can be activated by discrete dots.[40] So the neurons of V1 are not precluded from providing global location information to a shape filter. The contribution of details about contour orientation, curvature, and continuity may be useful for discriminating distinctive local contour features, making it possible to distinguish among similar objects. But these mechanisms would likely pivot off

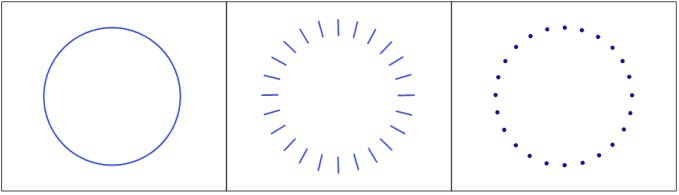

FIGURE 10. Each of the configurations is perceived as being circular based on the spatial location of the components. The attributes of orientation, curvature, and continuity that are present in the first panel have been modified or are absent in the second and third panels, yet these configurations also manifest circular shapes. An effective shape filter must see them as such.

the mechanisms that provide a global shape summary.

A global shape filter must register the locations of contour markers across visual space and summarize relationships irrespective of span differentials. One might consider concepts from algebraic topology, e.g., a Riemannian manifold, as might be needed for conformal mapping. However, this would require addresses within a coordinate system, and we have no evidence that an elementary contrast filter can generate address values to specify where it is located. An array of neurons can connect with great precision to another array as evidenced by the connections from retina to primary visual cortex. This is accomplished by chemical gradients and contact-tags[41] and one might hold out the possibility that synaptic transmitter chemistry could be used to specify coordinate addresses. If so, providing a silicon retina with the ability to deliver addresses from activated pixels provides at least the starting point for implementing global shape filters.

## VII. Converting 2D into 1D

I would like to consider some different possibilities. Perhaps neuronal mechanisms in the retina provide a means to convert the two-dimensional pattern of contour markers into a one-dimensional temporal message. I have previously suggested that the polyaxonal amacrine cells of the retina (PA1 neurons) generate spreading waves that might encode the relative positioning of markers.[42] The concept would be to turn spatial distances into temporal intervals. The spreading waves from each activated marker might converge at the centroid of the shape, with the arriving signals being converted into a temporal spike code. At least one neuromorphic shape encoding system has adopted this concept.[43]

Alternatively, in keeping with concepts advanced by Hopfield [44],[45] as well as Thorpe, VanRullen, and associates [46]-[49] a global shape-encoding mechanism might provide a sequential scan across marked locations, generating a spike from each as the scan-wave crosses. Where the scan-wave encountered a number of boundary markers that were aligned with its wave-front, simultaneous action potentials would be generated. The density of the action potentials being delivered by the optic nerve would vary according to the number of marked locations successively encountered by the scan wave. The two-dimensional boundary would thus be converted in a temporal code, wherein the density of spikes being generated at successive moments would reflect the shape, as sampled across the axis of a given scan-wave. Most shapes would require sampling scans in at least two directions to be reliably identified.

I have conducted two related experiments that provide some support for the feasibility of this shape-encoding concept.[50],[51] Both experiments used the inventory of novel (unknown) shapes described above. Each was sampled by vertical and horizontal scan-waves, as illustrated in Figure 11, providing raw histograms followed by a summary histogram that reflected the density of boundary markers encountered as the waves passed across the shape. Similarity of the summary histograms was determined for each pair of shapes using a least-squared calculation, yielding a similarity score for each pair. With an inventory of 480 shapes there were 114,960 pairs (combinations choose 2). These scores were ranked, providing a scale of similarity based on the degree of correspondence of the scan-generated histograms.

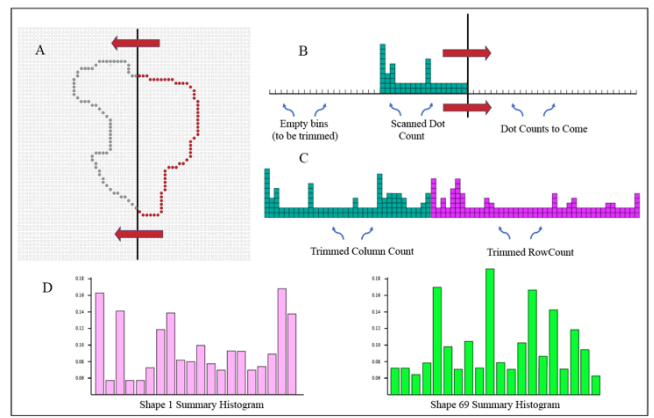

FIGURE 11. This method for identifying 2D shapes first creates a 1D summary histogram of each shape to be evaluated. A. For Shape1, a scan wave passes across the marked boundary locations, registering the number of markers encountered at any given moment. B. A raw histogram is constructed wherein the number of markers encountered in the scanned columns are plotted. C. The raw histograms for completed column and row scans are placed in tandem and trimmed to eliminate bins falling outside of the shape. D. The summary histogram on the left has re-binned and normalized the combined histogram (from C) to allow for comparison against other shape histograms. The summary histogram on the right is from a different shape. One can compare these histograms to determine the similarity of the shapes using a least squares match of bin counts. Comparison against an inventory of summary histograms can provide for shape identification.

Both experiments sampled pairs from across the range of similarity scores, then presented each pair in the match-recognition task. Pairs having high similarity scores were judged as being the "same" significantly more than those that had low similarity. Note that here, unlike the earlier studies using the match-recognition protocol, none of the comparison shapes were a low-density version of the target shape. One pair member was displayed as the target and the other pair member was displayed as the comparison shape. Nonetheless, when the score derived from the one-dimensional histograms indicated greater similarity of the pair members, the subjects were far more likely to judge the two shapes as being the same.

It may be noteworthy that one of these experiments also calculated a similarity index based on a well-established method for comparing two-dimensional boundaries.[51] This is the Procrustes index, wherein the shapes are sized and configured to have overlapping centroids, then boundary separations at corresponding locations around the boundary are measured and summed. The amount of separation as one passes around the boundary is meant to assess the degree to which the shapes are alike, and the minimum net span provides an index value for similarity. I can report that unlike the scan-based similarity scores discussed above, the Procrustes measures did not predict human judgments.

## VIII. Misleading Concepts

Trying to explain elemental memory storage by altering connectivity of successive large populations of neurons through multiple training trials has been a mistake. From the latter half of the 19th century and throughout the 20th, a vast majority of neuroscientists have adopted the view that the ability to learn and store new information is based on a change in the strength, number, or location of synaptic connections. Donald Hebb is best known for advancing this view [52], but it was a cornerstone of thinking across all sectors – including by those studying animal and human memory skills. When presented with the puzzle of how the brain could store new information, they speculated that one would need to alter strength or connectivity of the synapses. It was a guess, one that became an accepted principle well before there was any evidence that experience could produce synaptic changes.

Behavioral scientists reinforced this assumption by insisting that learning a new relationship between a stimulus and a response was achieved by trial and error. If a given behavior didn't lead to a beneficial outcome, the animal would try something else, and then something else again until an effective response was hit upon. The seeds of this concept can be traced back to Thorndike, who tested the ability of cats to escape from puzzle-boxes.[53] However, it fit well with the zeitgeist of American Psychology that wished to deny instinctive behaviors, especially as they might constrain or determine human propensity and aptitude.

The assumptions coming out of neuroscience and behavioral science created the dominant theories of how we learn new information, tailored here and there depending on the specifics to be learned. It was thought that additional experiences could provide the needed feedback to determine whether an initial change was beneficial or at least could move the behavior in the right direction. To explain how a single experience could produce a lasting change in memory, theorists invoked the concept of "reverberation,"[52] essentially a recirculation of neural activation that somehow could stand in for new experiences. None of the concepts were articulated with any specificity or precision, and all were vapid.

American psychologists never warmed to the methods used by European ethologists, where observation of the natural behavior of animals provided abundant evidence of inborn perceptual and cognitive skills. And in spite of increasing evidence that genetic information could provide for precise wiring both within a population of neurons and among populations,[55]-[57] neuroscientists continued to advocate for trial and error tailoring of synaptic strength or connectivity.

Modification of synaptic connectivity does occur as part of the maturational process,[55] as an adaptation to trauma,[58] and from interaction with the environment.[59]-[65] However, these changes generally take place slowly after extensive environmental exposure or intensive training. For the filters that can encode global attributes of a shape, it is likely that the design has already been formulated, having been tailored from the hundreds of millions of years of evolutionary development. While many mammals take a substantial amount of time for their brains to reach maturity, e.g., humans, it is a mistake to assume the eventual design of the shape-recognition filters came about through trial-and-error instruction.

## IX. Coda

A large part of the present message was to convey how neuroscience and behavioral science have provided misleading concepts for how the brain works. Those concepts are seldom conveyed with any specificity. Worse, many in the cognitive and brain sciences still explicitly or tacitly embrace the magic provided by conscious experience and free will.

Given the misdirection that was provided, I am amazed at the degree to which intelligent and creative electronics engineers, computer scientists, and the broader artificial intelligence community have succeeded in getting the ill-conceived neuroscience principles to work, or at least almost work. Real-world demands have been addressed with much greater rigor than has been true for experiments done by cognitive and brain researchers. The practical emphasis has clarified what kinds of mechanisms would be needed for a wide range of tasks.

Further, it is possible or even likely that the slow adjustment of connectivity through repeated encounters with the environment does provide for recovery of function after disease or injury has damaged normal connections of the brain. Therefore, the extensive work that has been done to develop effective neural networks may well serve as useful models for this recovery, and may inform how best to speed the recovery. Similar points could be made with respect to maturation of brain systems and the slow development of skills through practice.

Efforts to achieve more effective information processing will precede on a number of fronts. The major goal here has been to encourage those who are working to develop neuromorphic shape encoding filters. I do not think one must begin with orientation-selective filters, or use large populations of processing elements, or require a number of successive layers. It seems unwise to start with random connections that become functional across many training trials. Also, to focus too much on the human brain serves only to fog one's thinking about what is fundamental for the skill.

Figure 12 reminds us that modern fish can navigate a very complex seafloor, interact appropriately with other sea-life, and identify their own species based on complex cues. Each does so with a visual system consisting of retina and optic tectum. Even if human visual abilities can exceed theirs, they are most likely built on the basic encoding principles that these fish are using.

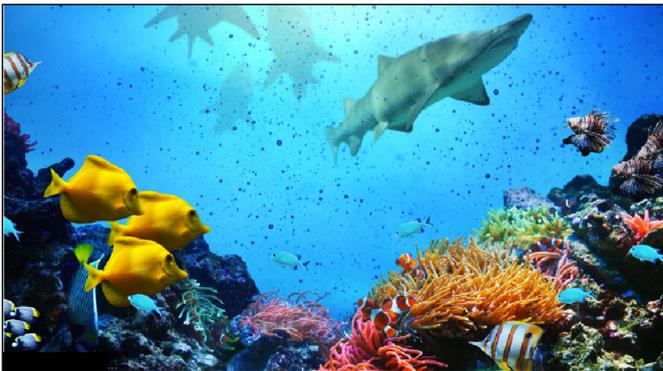

FIGURE 12. Ethologists have documented that fish are able to identify their own species based on complex visual patterns and quickly learn which shapes are predators and which are prey. The shape recognition is accomplished without benefit of cortex. It would be useful to develop neuromorphic circuits that could match the visual skills of fish.

## ACKNOWLEDGMENT
Funding for the Laboratory for Neuromeric Research has been provided by the Neuropsychology Foundation and the Quest for Truth Foundation